\documentclass[pdflatex,sn-basic]{sn-jnl}
\usepackage{amsmath,amssymb,amsfonts,amsthm,mathrsfs}
\usepackage{graphicx,xcolor,textcomp}
\usepackage{bm,booktabs,multirow,array,enumitem}
\usepackage{algorithm,algpseudocode,float}
\usepackage[title]{appendix}
\hypersetup{hidelinks,pdftitle={Exact conditional calibration of residual processes in linear mixed models},pdfauthor={Jakob Peterlin}}
\graphicspath{{figures/}}

\theoremstyle{thmstyleone}
\newtheorem{theorem}{Theorem}
\newtheorem{lemma}[theorem]{Lemma}
\newtheorem{proposition}[theorem]{Proposition}
\newtheorem{corollary}[theorem]{Corollary}
\theoremstyle{thmstyletwo}

\newcommand{\R}{\mathbb{R}}

\DeclareMathOperator{\col}{col}

\newcommand{\lab}{\mathcal{A}}

\DeclareMathOperator{\rank}{rank}

\newcommand{\figalt}[1]{}
\input{labels-supplement.tex}

\begin{document}

\title[Exact conditional calibration for linear mixed models]{Exact conditional calibration of residual processes in linear mixed models}

\author*[1]{\fnm{Jakob} \sur{Peterlin}}\email{jakob.peterlin@mf.uni-lj.si}
\affil*[1]{\orgdiv{Institute for Biostatistics and Medical Informatics}, \orgname{Faculty of Medicine, University of Ljubljana}, \orgaddress{\street{Vrazov trg 2}, \city{Ljubljana}, \postcode{1000}, \country{Slovenia}}}

\abstract{A cumulative residual test checks a fitted linear mixed model by summing residuals in the order of a fitted prediction and comparing their path with simulated paths. Large-sample calibrations can fail when small samples, short series or many fixed effects leave few residual degrees of freedom. We give a calibration that is exact at every sample size. Under independent homoscedastic Gaussian errors, the residual projected off all fixed and random design columns has a uniform direction, given its length and the response component inside the design span, and (restricted) maximum likelihood fits depend on the response only through these two summaries. Rotating the projected residual reproduces the conditional law of the observed path. The test is exact for nested or crossed factors, few groups, singular covariance estimates and any random-effect distribution, and one set of rotations calibrates a joint test over orderings and smoothing scales. The whitened process is covered once every rotation retains its fitted offset. Departures inside that span are invisible; outside it, a finite-sample bound gives power. In simulations with 6 to 1261 residual degrees of freedom the exact test held its 5\% level. Gaussian draws rejected 0.5\% of null samples with 6 and 4\% with 25, and up to 36\% when many fixed effects biased a maximum likelihood variance estimate; below 50 they lost 1 to 6 percentage points of power. In two small classical studies, of gasoline yield and rat growth, exact $p$-values were two to eight times smaller than with Gaussian draws.}



\maketitle

\section{Introduction}
\label{sec:intro}

Linear mixed models represent repeated measurements, nested populations
and crossed experimental factors through fixed and random effects
\citep{Laird82,Pinheiro2000,GelmanHill2007,Baayen2008,lme4}. A cumulative
residual process checks a fitted model by adding up residuals in the order
of a fitted prediction and comparing the resulting path with simulated
paths \citep{Su91,stute97,Lin02,Christensen10,Pan05}. We call the process
computed from the data the \emph{primary process} and the simulated ones
\emph{secondary processes}; the \emph{oracle process} is the same
construction at specified model parameters. The ordering decides what a
path can show: the fixed prediction, the full conditional prediction, or
the fixed prediction plus the predicted effects of one grouping factor.

\citet{LMMpaper} developed such tests for one grouping factor, with
secondary processes that perturb the residuals of independent groups. For
nested or crossed factors, the companion paper
\citep{asymptoticcalibration} projects the residuals off all fixed and
random design columns and shows that secondary processes from Gaussian
draws are valid as the residual dimension $r$, the number of observations
minus the rank of the design, grows. Calibration by Gaussian draws is
approximate in two ways. A Gaussian draw has a random length, which makes
the test conservative when $r$ is small. And when the marks are scaled by
the likelihood estimate of the error variance, that estimate can differ
from the residual mean square. In the simulations of
Section~\ref{sec:sim-small-r}, Gaussian draws reject 0.5\% of null samples
at the 5\% level when $r=6$ and about 4\% when $r=25$, and at $r=100$ they
are still 0.4 percentage points below the exact test; with twenty fixed
effects and an ML variance estimate they reject 36\% of the samples in 40
observations and 11\% in 100. The residual dimension can be much smaller
than the sample: 150 observations in 50 series of three, with random
intercepts and slopes, leave $r=50$. Mixed models with few observations
per group, many fixed effects or a rich random structure are of this kind.

This paper gives a calibration that is exact at every sample size. Let
$\bm R$ project off all fitted design columns and $\bm\Pi=\bm I-\bm R$.
Under Gaussian observation errors the direction of $\bm R\bm y$ is uniform
on the unit sphere of the residual space and independent of $\bm\Pi\bm y$
and $\|\bm R\bm y\|$. The companion paper uses this fact as a step in its
large-sample argument \citep[Proposition~7]{asymptoticcalibration}. Here it
becomes a test. A Gaussian maximum likelihood (ML) or restricted maximum
likelihood (REML) fit depends on the response only through $\bm\Pi\bm y$
and $\|\bm R\bm y\|$, so conditioning on these two summaries fixes the fit,
the orderings and the smoothing scales. Replacing the direction of
$\bm R\bm y$ by fresh uniform directions, which we call \emph{rotations},
then reproduces the conditional law of the primary process. This is a
conditional Monte Carlo test \citep{Engen1997,Lindqvist2005} and a
rotation test in the sense of \citet{langsrud2005}, applied to a fitted
mixed model. Our contributions are the following.
\begin{enumerate}[label=(\roman*),nosep]
\item \emph{An exact test} (Section~\ref{sec:exact}). For every projected
process the Monte Carlo $p$-value has its nominal conditional and
unconditional level at every sample size, for nested or crossed factors,
few groups, singular fitted covariances and any distribution of the random
effects (Theorem~\ref{thm:exact-main}).
\item \emph{A joint test.} One set of rotations, shared by all orderings,
statistics and smoothing scales, gives a single level-$\alpha$ test of the
model (Corollary~\ref{cor:joint-main}). The individual orderings then show
where a departure lies.
\item \emph{The whitened process.} The whitened residual process ordered
by the fixed prediction keeps a fitted offset inside the design span. Its
exact test, $F_E$, retains that offset in every rotation
(Proposition~\ref{prop:exact-main}), and the offset changes the power
(Section~\ref{sec:fixed}).
\item \emph{Scope and power} (Section~\ref{sec:power}). Conditioning makes
the tests blind to mean and covariance departures inside the fitted design
span, where they remain exactly valid (Corollary~\ref{cor:blind-main}).
Outside it, a finite-sample bound gives power once $r$ times the squared
excess of the observed statistic is large
(Theorem~\ref{thm:consistency-main}).
\end{enumerate}
Section~\ref{sec:simulations} reports simulations in five designs and a
study that follows the calibrations from $r=6$ upwards, and
Section~\ref{sec:application} applies the tests to two small studies, of
gasoline yield and of rat growth.

\section{Model, processes and tests}
\label{sec:model}

For a set $S$ of grouping factors, the projected process $O_S$ uses the
same residual marks and changes the ordering. The whitened process $F$
uses the fixed prediction and a different set of marks. We give both
definitions to specify which parts of their paths the conditioning holds
fixed.

\subsection{The model and the oracle processes}
The design $\mathcal D_n$ records the covariates and group memberships;
we treat these as given. Conditionally on $\mathcal D_n$, suppose
\begin{equation}
 \bm y=\bm X\bm\beta+\sum_{l=1}^L\bm Z_l\bm b_l+\bm\epsilon,
 \qquad \bm b_l\sim N(\bm0,\sigma^2\bm\Lambda_l),\qquad
 \bm\epsilon\sim N(\bm0,\sigma^2\bm I),
 \label{eq:model-main}
\end{equation}
with independent terms, $\sigma>0$ and $\bm\Lambda_l=\bm I_{m_l}\otimes\bm
D_l$, $\bm D_l\succeq0$. Here $\bm X\in\R^{n\times p}$ has full column rank,
factor $l$ has $m_l$ groups and $k_l$ random coefficients per group, and
$\bm Z_l\in\R^{n\times m_lk_l}$ is its design. Conditionally on
$\mathcal D_n$, the random-coefficient vectors of distinct groups are
independent within each factor; the vectors of different factors and the
observation errors are also mutually independent. Correlation among the
$k_l$ coefficients within a group is allowed through $\bm D_l$.
Here $\bm D_l\succeq0$ means positive semidefinite:
$\bm u^{\mathsf T}\bm D_l\bm u\ge0$ for every vector $\bm u$, allowing
zero variance components. More generally, $\bm A\preceq\bm B$ means
that $\bm B-\bm A$ is positive semidefinite; it compares quadratic forms,
not individual matrix entries.

This Gaussian specification defines the working likelihood. The exact
conditional test in Section~\ref{sec:exact} only requires Gaussian
observation errors independent of the component in $\col[\bm X,\bm Z]$;
the random-effect law in that component need not be Gaussian.
The numbers $L$, $p$ and
$k_1,\ldots,k_L$ are fixed, the group counts may grow with $n$ or stay fixed,
and the factors may be nested or crossed in any way (condition (M1) of
Section~\ref{S-sec:si-conditions}). Let
$\bm Z=[\bm Z_1,\ldots,\bm Z_L]$ and, for a set $S$ of factors,
\begin{equation}
 \bm R=\bm I-[\bm X,\bm Z][\bm X,\bm Z]^+,\quad
 \bm C_S=\sum_{l\in S}\bm Z_l\bm\Lambda_l\bm Z_l^{\mathsf T},\quad
 \bm V=\bm I+\bm C_{\{1,\ldots,L\}},\quad
 \bm H_S=\bm C_S\bm V^{-1},
 \label{eq:matrices-main}
\end{equation}
with $\bm C_\varnothing=\bm0$ and $\bm H_\varnothing=\bm0$. The superscript $+$ denotes the Moore--Penrose inverse, $\|\cdot\|$ is the
Euclidean norm, and
$\operatorname{col}[\bm X,\bm Z]$ is the space of all linear combinations
of the fitted design columns. Multiplication by $\bm R$ removes every
response pattern in this space; $\bm R$ is determined by the design,
without estimating variance components. The other matrices describe the
working covariance and prediction: $\sigma^2\bm C_S$ is the covariance
contribution of the factors in $S$, $\sigma^2\bm V$ is the full response
covariance, and $\bm H_S$ maps residuals to their predicted random-effect
contribution.

In particular, $\bm R$ is the orthogonal projection onto
$\operatorname{col}[\bm X,\bm Z]^\perp$ whatever the ranks of $\bm X$ and
$\bm Z$, and $\bm R\bm X=\bm0$, $\bm R\bm Z=\bm0$, $\bm R\bm V\bm R=\bm R$
and $\bm R\bm H_S=\bm0$. With $\bm e=\bm y-\bm X\bm\beta$, $\bm H_S\bm e$ is
the contribution of the factors in $S$ to the best linear unbiased prediction
\citep{Robinson91}, and $\bm R\bm e=\bm R\bm\epsilon$ is free of the fixed
effects and of every random effect. The projection $\bm P=\bm I-\bm Z\bm Z^+$
off the random-effect design alone, which \citet{LMMpaper} used, is related
to the least confounded residuals of \citet{HildenMinton1995} and
\citet{Nobre2007}; $\bm R\bm y=\bm P\bm y$ when
every fixed regressor is also a random regressor for some factor. In general
$\bm R$ removes in addition the component of $\bm P\bm y$ in
$\operatorname{col}(\bm P\bm X)$, at most $p$ dimensions;
Section~\ref{sec:exact} explains why this is the right choice for the primary
processes.

A \emph{mark} is the residual contribution of an observation to the
cumulative sum; its \emph{ordering value} determines when it enters that
sum. A process is labelled by
$a\in\lab=\{F,O,O_1,\ldots,O_L,O_\varnothing\}$ and defined by an ordering
vector $\bm y^a$, a residual map $\bm B^a$ and marks
$\bm e^a=\sigma^{-1}\bm B^a\bm e$. The whitened process $F$ orders by the
fixed prediction $\bm y^F=\bm X\bm\beta$ and whitens,
$\bm B^F=\bm V^{-1/2}$. The projected process $O_S$ orders by the fixed
prediction plus the predicted random effects of the factors in $S$ and
projects off the fixed and every random-effect design,
\begin{equation}
 \bm y^{O_S}=\bm X\bm\beta+\bm H_S\bm e,\qquad \bm B^{O_S}=\bm R.
 \label{eq:orderings-main}
\end{equation}
We write $O=O_{\{1,\ldots,L\}}$ and $O_l=O_{\{l\}}$; the
\emph{fixed-prediction process} $O_\varnothing$ orders by
$\bm X\bm\beta$ alone. All projected processes share their marks and
differ only in their ordering. For example, with subjects and items,
$O_{\rm subj}$ orders these marks by the fixed prediction plus the
predicted subject contribution, while $O_{\rm item}$ uses the item
contribution. Table~\ref{tab:notation} collects the processes and the
test labels.

For a smoothing scale $\lambda\in(0,\infty]$, let $\sigma_\lambda$ be a
nonincreasing weight with values in $[0,1]$. Indicator weights use
$\sigma_\infty(x)=1\{x\le0\}$; the smooth weights for finite $\lambda$ are
specified in Section~\ref{sec:scale}. Write
$\bm\sigma_\lambda(\bm v,t)=(\sigma_\lambda(v_i-t))_{i=1}^n$.
The oracle process is
\begin{equation}
 W_n^a(t)=n^{-1/2}\bm\sigma_\lambda(\bm y^a,t)^{\mathsf T}\bm e^a.
 \label{eq:processes-main}
\end{equation}
It is a theoretical benchmark: its parameters are supplied rather than
estimated from the sample, although its residuals and predictions still
depend on the response. Under the null we use the true model parameters.
For indicator weights, the oracle process is
$n^{-1/2}\sum_{i:y_i^a\le t}e_i^a$:
the marks are accumulated in the order given by $\bm y^a$. Finite
$\lambda$ smooths the inclusion of each observation.

\subsection{The primary and secondary processes}
Replacing the oracle parameters by their fitted values gives the
primary process, the path actually computed from the observed data:
\begin{equation}
 \widehat W_n^a(t)=n^{-1/2}\bm\sigma_\lambda(\widehat{\bm y}^a,t)^{\mathsf T}
 \widehat{\bm e}^{a},\qquad
 \widehat{\bm e}^{a}=\widehat\sigma^{-1}\widehat{\bm B}^a
 (\bm y-\bm X\widehat{\bm\beta}),
 \label{eq:fitted-process-main}
\end{equation}
where hats denote ML estimates, or REML covariance estimates with their
generalised least squares (GLS) fixed effects. Write
$\widehat{\bm e}=\bm y-\bm X\widehat{\bm\beta}$. The predicted random-effect
vector of factor $l$ is
\[
 \widehat{\bm b}_l=\widehat{\bm\Lambda}_l\bm Z_l^{\mathsf T}
 \widehat{\bm V}^{-1}\widehat{\bm e},\qquad l=1,\ldots,L.
\]
The corresponding orderings are
$\widehat{\bm y}^{F}=\bm X\widehat{\bm\beta}$ and
$\widehat{\bm y}^{O_S}=\bm X\widehat{\bm\beta}
+\sum_{l\in S}\bm Z_l\widehat{\bm b}_l$.
Since $\bm R\bm X=\bm0$, the fitted marks of every
projected process are $\widehat\sigma^{-1}\bm R\bm y$.

The secondary processes are \emph{rotations} of the primary marks: new
residual directions with the observed residual length. Set
\[
 \bm z_0=\widehat\sigma^{-1}\bm R\bm y,\qquad
 \bm c_a=\begin{cases}
  \bm0,&a\text{ projected},\\
  \widehat{\bm e}^{F}-\bm z_0,&a=F.
 \end{cases}
\]
For $b=1,\ldots,B$, independently draw $\bm\xi_b\sim N(\bm0,\bm I)$ and
define
\begin{equation}
 \bm z_b=\|\bm z_0\|\frac{\bm R\bm\xi_b}{\|\bm R\bm\xi_b\|},\qquad
 \bm e_b^{a,\#}=\bm c_a+\bm z_b,\qquad
 W_{n,b}^{a,\#}(t)=n^{-1/2}
 \bm\sigma_\lambda(\widehat{\bm y}^a,t)^{\mathsf T}\bm e_b^{a,\#}.
 \label{eq:secondary-exact-main}
\end{equation}
Here $\rank(\bm R)>0$; otherwise the projected marks vanish and there is
no projected diagnostic. This constructs each secondary process in three
steps: project a Gaussian vector into the residual space, divide by its
own length to obtain a random unit direction, and multiply by the
primary residual length $\|\bm z_0\|$. The two lengths come from different
vectors: the normalisation removes the simulated length and retains the
observed one. Accumulating the resulting marks then produces a secondary
path. For $F$, the fitted offset $\bm c_F$ is added back to every draw;
Section~\ref{sec:conditional-law} explains why this component stays fixed.
Thus the primary and secondary processes use
the same accumulation rule, fitted ordering, grid and smoothing scale;
only the residual direction changes. No refitting is needed. The same
$\bm z_b$ is used for every process and scale within draw $b$.
We write $O_E$ for the exact test of a projected process and $F_E$ for
that of $F$. Section~\ref{sec:exact} proves their conditional validity.
The companion paper scales the projected marks by the residual standard
deviation $\|\bm R\bm y\|/\sqrt r$ instead of $\widehat\sigma$. For the
projected processes this changes no rank under rotations, because a
common positive factor cancels.

\begin{table}[!hb]
\caption{Exact tests and the quantities defining their primary and secondary
processes. Write
$\widehat{\bm e}=\bm y-\bm X\widehat{\bm\beta}$. Hats denote ML estimates, or
REML covariance estimates with their GLS fixed effects.}
\label{tab:notation}
\centering\small
\renewcommand{\arraystretch}{1.2}
\begin{tabular}{@{}
 >{\raggedright\arraybackslash}p{.15\textwidth}@{\hspace{.02\textwidth}}
 >{\raggedright\arraybackslash}p{.21\textwidth}@{\hspace{.02\textwidth}}
 >{\raggedright\arraybackslash}p{.17\textwidth}@{\hspace{.02\textwidth}}
 >{\raggedright\arraybackslash}p{.41\textwidth}@{}}
\toprule
Test & $\widehat{\bm y}^a$ & $\widehat{\bm e}^a$ & $\bm e_b^{a,\#}$\\
\midrule
$O_E$\newline Overall
 & $\bm X\widehat{\bm\beta}+\sum_{k=1}^L\bm Z_k\widehat{\bm b}_k$
 & $\widehat\sigma^{-1}\bm R\bm y$
 & $\displaystyle\frac{\|\bm R\bm y\|}{\widehat\sigma}
       \frac{\bm R\bm\xi_b}{\|\bm R\bm\xi_b\|}$\\
\addlinespace[.6em]
$O_{l,E}$\newline Factor $l$
 & $\bm X\widehat{\bm\beta}+\bm Z_l\widehat{\bm b}_l$
 & $\widehat\sigma^{-1}\bm R\bm y$
 & as for $O_E$\\
\addlinespace[.6em]
$O_{\varnothing,E}$\newline Fixed prediction
 & $\bm X\widehat{\bm\beta}$
 & $\widehat\sigma^{-1}\bm R\bm y$
 & as for $O_E$\\
\addlinespace[.6em]
$F_E$\newline Whitened
 & $\bm X\widehat{\bm\beta}$
 & $\widehat\sigma^{-1}\widehat{\bm V}^{-1/2}\widehat{\bm e}$
 & $\displaystyle\widehat\sigma^{-1}(\widehat{\bm V}^{-1/2}-\bm R)\widehat{\bm e}
     +\frac{\|\bm R\bm y\|}{\widehat\sigma}
       \frac{\bm R\bm\xi_b}{\|\bm R\bm\xi_b\|}$\\
\bottomrule
\end{tabular}
\par\smallskip
\begin{minipage}{\textwidth}\footnotesize
Draw $\bm\xi_b\sim N(\bm0,\bm I)$ independently across
$b$, using the same draw for all processes and scales. Every secondary
process keeps the fitted ordering fixed; the first term in
$\bm e_b^{F,\#}$ is the fitted offset $\bm c_F$.
\end{minipage}
\end{table}

\subsection{The smoothing scale}
\label{sec:scale}
For finite $\lambda$, we use the quintic weight of \citet{LMMpaper}:
$\sigma_\lambda(x)=1$ for $x\le0$,
$1-10(\lambda x)^3+15(\lambda x)^4-6(\lambda x)^5$ for $0<x<1/\lambda$,
and $0$ for $x\ge1/\lambda$. Thus the transition from inclusion to
exclusion occurs over an interval of width $1/\lambda$; indicator weights
are recovered as $\lambda\to\infty$. The proofs use only that the weights
are nonincreasing with values in $[0,1]$.
We use $\lambda=c/I_n$, where $I_n$ is the fitted 95th minus 5th
percentile and $c\in\{1,2,10,100,1000\}$. Indicator weights give a
separate column. The fallback for a zero range uses only the fit
(Section~\ref{S-sec:si-simulations}).

\subsection{Statistics and $p$-values}
\label{sec:pvalues}
\label{sec:procedure}
The Kolmogorov--Smirnov (KS) and root
Cram\'er--von Mises (CvM) statistics evaluate a path at the fitted ordering
values,
\begin{equation}
 T_{{\rm KS},n}(W)=\max_i|W(\widehat y_i^a)|,\qquad
 T_{{\rm CvM},n}(W)=\Bigl\{n^{-1}\textstyle\sum_iW(\widehat y_i^a)^2\Bigr\}^{1/2},
 \label{eq:statistics-main}
\end{equation}
Tied ordering values enter together, with their observed multiplicities.

The Monte Carlo $p$-value of the primary statistic $T_0$ among the secondary
statistics $T_1^\#,\ldots,T_B^\#$ is
\begin{equation}
 \widehat p=\Bigl\{1+\textstyle\sum_{b=1}^B1\{T_b^\#\ge T_0\}\Bigr\}/(B+1).
 \label{eq:mc-main}
\end{equation}
We report root CvM as the primary statistic.

For selection over scales, we take the smallest marginal $p$-value and
calibrate that minimum with the same draws,
the device of \citet{WestfallYoung1993} applied to a grid of scales as in
\citet{Horowitz2001}: with $T_{bc}$ the statistic at scale $c$ of row $b$,
$b=0$ the observed one,
\begin{equation}
 p_{bc}=\frac{\sum_{r=0}^B1\{T_{rc}\ge T_{bc}\}}{B+1},\qquad
 M_b=\min_{c\in\mathcal C}p_{bc},\qquad
 \widehat p_{\min}=\frac{\sum_{b=0}^B1\{M_b\le M_0\}}{B+1}.
 \label{eq:minp-main}
\end{equation}
Use one rotation across all the prespecified columns of a row.
Proposition~\ref{prop:finite-rank-main} gives the rank bounds.

\begin{algorithm}[H]
\caption{Exact conditional residual-process test}\label{alg:exact}
\begin{algorithmic}[1]
\State Fit the model by ML or REML and compute $\bm R$, $r=\rank\bm R>0$,
$\widehat{\bm y}^a$ and the prespecified statistic columns.
\State Set $\bm z_0=\widehat\sigma^{-1}\bm R\bm y$;
$\bm c_a=0$ for projected processes and
$\bm c_F=\widehat{\bm e}^{F}-\bm z_0$ for $F_E$.
\For{$b=1,\ldots,B$}
\State Draw $\bm\xi_b\sim N(\bm0,\bm I)$ and set
$\bm z_b=\|\bm z_0\|\bm R\bm\xi_b/\|\bm R\bm\xi_b\|$.
\EndFor
\State Evaluate every row $b=0,\ldots,B$ using marks $\bm c_a+\bm z_b$
and the same fitted ordering, grid and scales.
\State Apply \eqref{eq:minp-main} to the combined columns for one global
$p$-value; retain individual paths for interpretation.
\end{algorithmic}
\end{algorithm}

Report the residual dimension $r$, the fitted design and error
assumptions, the global $p$-value and the individual paths. Pointwise
bands or collections of secondary paths illustrate the conditional
law; they do not define an additional test.

The construction is implemented using \texttt{GOFLMM}'s
\texttt{LMM\_P1} projection with \texttt{whitening=:symmetric} and the
shared-draw routines \texttt{scripts/exact1\_lib.jl}. The main costs are
factorising the fitted design, projecting the Gaussian draws, and the
whitening used by $F_E$. Neither the covariance nor the fixed effects
need to be refitted for a rotation. The structured whitener is a faster
alternative but depends on group labelling; the symmetric root is the
default choice (Section~\ref{S-sec:si-simulations}).

\section{Exact conditional calibration}
\label{sec:exact}

We now justify the primary and secondary processes defined in
Section~\ref{sec:model}. The argument applies to both the projected tests
and $F_E$: fitting fixes the orderings, the residual length and the
$F$ offset, while the residual direction remains uniform. We identify
this conditioning information, establish the conditional law, and then
prove the validity of the individual and joint tests.

\subsection{Conditioning information and fitting}
\label{sec:exact-fitted}
Let $\bm\Pi=\bm I-\bm R$ be the orthogonal projection onto
$\operatorname{col}[\bm X,\bm Z]$ and $r=\rank(\bm R)$. A fit is feasible
if its covariance has the form
$\widehat{\bm V}=\bm I+\bm Z\widehat{\bm G}\bm Z^{\mathsf T}$ with
$\widehat{\bm G}\succeq0$ and $\widehat\sigma>0$, which constrained ML and
REML fits are by construction. Write
$\bm\Gamma=(\bm X^{\mathsf T}\widehat{\bm V}^{-1}\bm X)^{-1}
\bm X^{\mathsf T}\widehat{\bm V}^{-1}$ and $\bm M=\bm I-\bm X\bm\Gamma$ for
the GLS coefficient and residual maps and $\widehat{\bm H}_S$ for the
prediction matrices of \eqref{eq:matrices-main} at the fitted covariance,
so that the primary marks are
$\widehat\sigma^{-1}\widehat{\bm V}^{-1/2}\bm M\bm y$ for $F$ and
$\widehat\sigma^{-1}\bm R\bm M\bm y$ for every projected process. Everything
rests on the following identities. They express that the covariance
parameters act within the random-effect design span, while the relative
covariance on its orthogonal complement is the identity. The likelihood
therefore uses the residual length but cannot distinguish residual
directions of the same length.

\begin{lemma}[Residual-space identities]\label{lem:residual-space}
For every feasible covariance $\widehat{\bm V}$ and every $S$,
$\widehat{\bm V}^{\pm1}\bm R=\bm R\widehat{\bm V}^{\pm1}=\bm R$ and
$\widehat{\bm V}^{\pm1/2}\bm R=\bm R\widehat{\bm V}^{\pm1/2}=\bm R$;
$\bm\Gamma\bm R=\bm0$, $\bm M\bm R=\bm R\bm M=\bm R$ and
$\widehat{\bm H}_S\bm R=\bm0$; $\bm A\bm R=\bm R\bm A=\bm R$ for every
$\bm A=\bm I+\bm Z\bm\Theta\bm Z^{\mathsf T}$; and, for every $\bm\gamma$,
\begin{equation}
 (\bm y-\bm X\bm\gamma)^{\mathsf T}\widehat{\bm V}^{-1}(\bm y-\bm X\bm\gamma)
 =(\bm\Pi\bm y-\bm X\bm\gamma)^{\mathsf T}\widehat{\bm V}^{-1}
  (\bm\Pi\bm y-\bm X\bm\gamma)+\|\bm R\bm y\|^2,
 \label{eq:likelihood-split}
\end{equation}
so that the ML and the REML objectives depend on $\bm y$ only through
$\bm\Pi\bm y$ and $\|\bm R\bm y\|^2$.
\end{lemma}
\begin{proof}
Since $\bm Z^{\mathsf T}\bm R=0$, the fitted covariance is block diagonal
on $\operatorname{ran}(\bm R)\oplus\operatorname{ran}(\bm\Pi)$ and equals
the identity on the first block. This proves the inverse and square-root
identities; $\bm X^{\mathsf T}\bm R=0$ gives the GLS identities. The two
summands in
$\bm y-\bm X\bm\gamma=(\bm\Pi\bm y-\bm X\bm\gamma)+\bm R\bm y$
are orthogonal also for the fitted precision, proving
\eqref{eq:likelihood-split}. The additional REML determinant depends only
on the covariance and the design.
\end{proof}

Identity \eqref{eq:likelihood-split} is the classical decomposition of the
generalised residual sum of squares into the part computed from the
mixed-model equations and the pure-error sum of squares
$\bm y^{\mathsf T}\bm R\bm y$ \citep{Searle1992}; the exact tests of variance
components of \citet{Seely1983} and \citet{Ofversten1993} condition on the
same quantity. What is used here is its consequence for the primary
processes: the ML or REML fit, the fitted orderings and the relative
scales are all functions of $(\bm\Pi\bm y,\|\bm R\bm y\|)$, which is what
the rotation argument below needs.

In particular the primary marks of every projected process are
$\widehat\sigma^{-1}\bm R\bm y$. Let
$\mathcal H=\sigma\{\bm\Pi\bm y,\|\bm R\bm y\|\}$, the fitted information; a
quantity is $\mathcal H$-measurable when it is a function of the fitted
information. Here $\sigma\{\cdot\}$ denotes the sigma-algebra generated by
the summaries. Conditioning on $\mathcal H$ restricts attention to
responses $\bm\Pi\bm y+\bm u$ with $\bm u\in\operatorname{ran}(\bm R)$
and $\|\bm u\|=\|\bm R\bm y\|$. All these responses give the same fit,
ordering values and relative scales; only their residual directions differ.

Throughout we assume $r>0$; at $r=0$ the projected marks vanish and the
implementation returns no projected diagnostic. We take the design as
fixed. With a random design, every statement holds conditionally on
$\mathcal D_n$, hence unconditionally, when $\bm\epsilon$ is independent
of the design and of $\bm b$.

Ties between equally good fits are broken by a measurable rule in the summaries
$(\bm\Pi\bm y,\|\bm R\bm y\|)$. A deterministic optimiser qualifies
whenever its starting values are functions of the summaries, as the
ordinary least squares residual variance is, since
$\|(\bm I-\bm X\bm X^+)\bm y\|^2=\|\bm R\bm y\|^2+\|(\bm\Pi-\bm X\bm X^+)\bm y\|^2$,
and as the default starting values of \texttt{MixedModels.jl} and
\texttt{lme4} are, being constants.
Nothing below needs the global minimiser: whatever the optimiser returns
qualifies, provided it is a function of the summaries.

In the statements below the model is written as
$\bm y=\bm X\bm\beta+\bm Z\bm b+\sigma\bm\epsilon$ with a standard Gaussian
$\bm\epsilon$, that is, $\bm\epsilon$ is $\sigma^{-1}$ times the error of
\eqref{eq:model-main}. The hypothesis tested by the rotation calibration is
\begin{equation}
 H_0^{E}:\quad \bm y=\bm X\bm\beta+\bm Z\bm b+\sigma\bm\epsilon,\qquad
 \bm\epsilon\sim N(\bm0,\bm I)\ \text{independent of }\bm b,
 \label{eq:H0-exact}
\end{equation}
with the fitted $\bm X$ and $\bm Z$ and an arbitrary law of $\bm b$:
the component of $\bm y$ orthogonal to $\col[\bm X,\bm Z]$ is $\sigma$
times a standard Gaussian vector, independent of the component in the
span. Only the spherical symmetry of $\bm\epsilon$ is used below, so the
Gaussian law could be replaced by any spherically symmetric law independent
of $\bm b$; for independent coordinates, however, spherical symmetry means
Gaussianity in dimension at least two.

\subsection{Conditional law of the primary processes}
\label{sec:conditional-law}

\begin{proposition}[Exact conditional law of the primary processes]
\label{prop:exact-main}
Under $H_0^E$ in \eqref{eq:H0-exact}, the Gaussian ML and REML fits,
their GLS coefficients, fitted orderings and relative scales are
$\mathcal H$-measurable. With $\bm z_0=\widehat\sigma^{-1}\bm R\bm y$
as in Section~\ref{sec:model}, the primary mark vector of every projected
process has, conditionally on $\mathcal H$, the law
\begin{equation}
 \|\bm z_0\|\,\frac{\bm R\bm\xi}{\|\bm R\bm\xi\|},\qquad
 \bm\xi\sim N(\bm0,\bm I)\ \text{independent of }\mathcal H.
 \label{eq:conditional-representation}
\end{equation}
For $F$ formed with an $\mathcal H$-measurable whitener $\bm A$ satisfying
$\bm A\widehat{\bm V}\bm A^{\mathsf T}=\bm I$ and $\bm A\bm R=\bm R$,
the primary mark vector has the conditional
law of \eqref{eq:conditional-representation} plus the fixed offset
\[
 \bm c_F=\widehat\sigma^{-1}\bm A\bm M\bm\Pi\bm y
        =\widehat{\bm e}^{F}-\bm z_0\in\operatorname{ran}(\bm\Pi),
\]
which is $\mathcal H$-measurable. In particular, this covers the symmetric
root used in Section~\ref{sec:model}.
\end{proposition}

Thus conditioning on $\mathcal H$ fixes the primary residual length while
leaving its direction uniform on the residual sphere. The appearance of
$\|\bm z_0\|$ in the conditional law is therefore deliberate: it is a
known radius under this conditioning, and the new vector $\bm\xi$
supplies only the random direction. Projecting off $\bm X$ as well as $\bm Z$
removes the fit-dependent offset present in the $\bm P$-projected residuals
of \citet{LMMpaper}, whose projection is taken off the random-effect design
alone. Table~\ref{S-tab:projection-compare} compares that construction with
the $\bm R$-projected residuals used here.

The offset is retained in every secondary $F$ path. It distinguishes
$F_E$ from the fixed-prediction process $O_\varnothing$, which uses the
same ordering but has no offset. Proposition~\ref{S-prop:exact-F}
gives the matrix derivation and the admissible structured whitener;
Section~\ref{sec:fixed} discusses the implications for detection.

\subsection{Validity of individual and joint tests}
\label{sec:exact-test}
We recall the exchangeable-row result of
\citet[Proposition~1]{asymptoticcalibration} to specify both boundaries
when ranks tie.

\begin{proposition}[Symmetric calibration over scales]
\label{prop:finite-rank-main}
Let $N=B+1$, $J=|\mathcal C|\ge1$ and $k_\alpha=\lfloor\alpha N\rfloor$ for
$0<\alpha<1$. If the $N$ rows $(T_{bc})_{c\in\mathcal C}$, $b=0,\ldots,B$,
are exchangeable, then $\Pr(\widehat p_{\min}\le\alpha)\le k_\alpha/N$. If,
in addition, the $N$ statistics at each scale are pairwise distinct almost
surely, then
$\max\{0,(k_\alpha+1-J)/N\}\le\Pr(\widehat p_{\min}\le\alpha)\le k_\alpha/N$.
No independence between scales is required.
\end{proposition}

With $B=999$ and five scales, the rejection probability at level $0.05$
lies in $[0.046,0.050]$ when the no-ties hypothesis holds. Otherwise only
the upper bound is guaranteed; an identically zero process has $p=1$.
Here ``exact'' refers to finite-sample control of the rejection probability;
ties and the discrete Monte Carlo ranks can make the test conservative.

\begin{theorem}[Exact conditional calibration]\label{thm:exact-main}
In the setting of Proposition~\ref{prop:exact-main}, fix a process $a$,
one of the two statistics and $J$ scale columns whose scales, weights
and grids are $\mathcal H$-measurable. This includes the relative scales
$c/I_n$ and fitted grids of Section~\ref{sec:model}. Generate the
secondary processes by \eqref{eq:secondary-exact-main}, using independent
Gaussian vectors $\bm\xi_b$ and keeping the fitted orderings fixed. Then,
conditionally on $\mathcal H$, the primary statistic vector and the $B$
secondary statistic vectors are independent and identically distributed.
Consequently, for every $n$, every fixed design, every distribution of the
random effects and every $B$, the single-column $p$-value \eqref{eq:mc-main}
satisfies $\Pr(\widehat p\le\alpha\mid\mathcal H)\le\lfloor\alpha(B+1)\rfloor/(B+1)$,
with equality when the statistics are almost surely distinct, and the
minimum-$p$ value \eqref{eq:minp-main} satisfies the bounds of
Proposition~\ref{prop:finite-rank-main} under its respective hypotheses,
conditionally on $\mathcal H$ and unconditionally.
\end{theorem}

\begin{proof}[Proof sketch]
By Lemma~\ref{lem:residual-space} both likelihood objectives, hence their
minimisers, the GLS coefficients, the orderings and the relative scales
depend on $\bm y$ only through $\bm\Pi\bm y$ and $\|\bm R\bm y\|$, and the
marks are $\bm c_a+\widehat\sigma^{-1}\bm R\bm y$
with $\bm R\bm y=\sigma\bm R\bm\epsilon$ independent of $\bm\Pi\bm y$. The
direction of a standard Gaussian vector in $\operatorname{ran}(\bm R)$ is
uniform on its unit sphere and independent of the length, which is
\eqref{eq:conditional-representation}; the $B+1$ mark vectors are therefore
conditionally independent and identically distributed given $\mathcal H$,
the rank of the primary statistic among $B+1$ exchangeable values is at most
$k$ with probability at most $k/(B+1)$, and the minimum-$p$ bounds are
Proposition~\ref{prop:finite-rank-main} applied conditionally
(Section~\ref{S-sec:si-exact}).
\end{proof}

\begin{corollary}[Joint calibration across processes and scales]
\label{cor:joint-main}
In Theorem~\ref{thm:exact-main}, let the columns index any finite,
prespecified family of process--scale--statistic combinations with
$\mathcal H$-measurable weights and grids. Use the same projected direction
in each column of a row, including the fixed offset for $F_E$. Then the
joint minimum-$p$ rule has conditional size at most
$\lfloor\alpha(B+1)\rfloor/(B+1)$. The lower bound of
Proposition~\ref{prop:finite-rank-main} applies with $J$ equal to the total
number of columns when its no-ties condition holds.
\end{corollary}
\begin{proof}
Each row is the same measurable vector-valued function of one independent
uniform residual direction. Theorem~\ref{thm:exact-main} therefore gives
exchangeable rows, to which Proposition~\ref{prop:finite-rank-main} applies.
\end{proof}
The corollary gives one test of the common model null, not a claim of
strong family-wise error control for separate component null hypotheses.
After a global rejection, a factor-specific path describes an ordering
along which the departure is visible; it does not identify a unique
misspecified term. Orderings by a design covariate are covered in the
same way. A search over orderings that
uses the residual direction must itself be calibrated and is not made
valid by the corollary.

For projected paths the common observed radius cancels from all ranks
by positive homogeneity. At small positive residual rank ties can make
the tests conservative. With the whitened process the offset remains in
every row. The construction uses a single fit and applies equally to a
singular fitted random-effect covariance.

\subsection{Comparison with Gaussian draws}
\label{sec:gaussian-approx}
Gaussian draws replace the rotation by a fresh Gaussian vector at the
fitted covariance. They generate
$\bm y_b^\#=\bm X\widehat{\bm\beta}+\widehat\sigma\widehat{\bm V}^{1/2}\bm\xi_b$
and re-estimate only the fixed effects:
\begin{equation}
 \widehat{\bm\beta}_b^\#=(\bm X^{\mathsf T}\widehat{\bm V}^{-1}\bm X)^{-1}
 \bm X^{\mathsf T}\widehat{\bm V}^{-1}\bm y_b^\#,\qquad
 \widehat W_{n,b}^{a\#}(t)=\frac{
 \bm\sigma_\lambda(\widehat{\bm y}^a,t)^{\mathsf T}\widehat{\bm B}^a
 (\bm y_b^\#-\bm X\widehat{\bm\beta}_b^\#)}{\widehat\sigma\sqrt n}.
 \label{eq:secondary-main}
\end{equation}
For a projected process the marks are $\bm R\bm\xi_b$, the Gaussian draws
of \citet{asymptoticcalibration}. A rotation rescales each of them by
$d\sqrt r/\|\bm R\bm\xi_b\|$, where
\[
 d=\frac{\|\bm z_0\|}{\sqrt r}
   =\frac{\|\bm R\bm y\|}{\widehat\sigma\sqrt r}
\]
is the observed radius relative to its Gaussian root mean square. The two
calibrations therefore differ only through this radial factor. As
$r\to\infty$, $\|\bm R\bm\xi_b\|/\sqrt r\to_p1$; with $d\to_p1$ and
suitable nondegeneracy they agree in the limit
\citep[Proposition~7]{asymptoticcalibration}. For finite $r$ they differ
in two ways. KS and root CvM are positively homogeneous in the marks, so
the statistic of a Gaussian draw is that of its rotation multiplied by
$\|\bm R\bm\xi_b\|/(d\sqrt r)$, and this factor is independent of the
rotation. First, the length of a Gaussian draw varies: when $d=1$ the
factor has mean square one and only spreads the reference distribution.
The companion paper divides the marks by the residual standard deviation
$\|\bm R\bm y\|/\sqrt r$ in place of $\widehat\sigma$, which sets
$d=1$, so that only this first effect remains. Second, $d$ differs from
one when
$\widehat\sigma^2$ is far from the residual mean square
$\|\bm R\bm y\|^2/r$: for an ML fit with many fixed effects $d>1$, and
the reference statistics are too small. Figure~\ref{fig:small-r} shows
both effects.

For $F$ the difference also concerns the component inside the fitted
design span: the exact test retains $\bm c_F$, whereas Gaussian draws
replace it by a random component, whose covariance is derived in
Proposition~\ref{S-prop:exact-F}.

\section{Scope and power}\label{sec:power}
The conditioning summaries determine which mean shifts remain visible
and how the whitened offset enters a power calculation.

\subsection{Departures inside the fitted span}
\label{sec:blindness}

\begin{corollary}[Exact validity under misspecification inside the fitted span]
\label{cor:blind-main}
Let $\bm y=\bm X\bm\beta+\bm v+\sigma\bm\epsilon$, where
$\bm v\in\operatorname{col}[\bm X,\bm Z]$ almost surely has an arbitrary
distribution and $\bm\epsilon\sim N(\bm0,\bm I)$ is independent of $\bm v$,
and fit any feasible covariance by a measurable function of the summaries
$(\bm\Pi\bm y,\|\bm R\bm y\|)$, in particular by ML or REML. Then the conclusions of
Theorem~\ref{thm:exact-main} hold for the projected processes: their exact
test is exactly valid although the law of $\bm v$ may violate every
covariance restriction of the fitted family and may shift the mean inside
$\operatorname{col}[\bm X,\bm Z]$.
\end{corollary}
\begin{proof}
$\bm R\bm y=\sigma\bm R\bm\epsilon$ whatever $\bm v$, and $\bm v$ enters
$\bm\Pi\bm y$ only; the argument of Theorem~\ref{thm:exact-main} applies
verbatim (Section~\ref{S-sec:si-exact}).
\end{proof}

For example, a covariate that is constant within schools is in the
span of fitted school intercepts. Conditioning removes that mean shift.
A covariance error outside the span can still change the residual law.

\subsection{The fitted offset}
\label{sec:fixed}
The fixed-prediction process $O_\varnothing$ and $F_E$ use the same
ordering $\bm X\widehat{\bm\beta}$ and the same rotations. Only $F_E$ adds
the offset $\bm c_F$ to every row. The offset is the whitened image of
the part of the response inside the design span, that is, of the fitted
mean and the predicted random effects, and it is fixed by the
conditioning. It carries no signal of its own: a departure inside the
span moves the offset path of the primary and of every secondary row
alike. It changes the ranks because KS and CvM measure distance from
zero. The path of the projected marks is then judged around the offset
path instead of around zero, and in the CvM statistic the offset enters
through its cross term with that path.

The offset therefore helps when the projected residual and the residual
inside the span are aligned along the fixed prediction. With an omitted
quadratic term and five or ten observations per subject, the fitted
subject slopes absorb part of the curvature, the two residuals are
aligned, and $F_E$ is more powerful than $O_\varnothing$. In the crossed
design each subject has about $\sqrt n$ observations, the projection
keeps almost all of the curvature, and the two tests have the same power.
Section~\ref{sec:sim-detection} gives the rates.

\subsection{Mean drift and power}
\label{sec:mean-power}

Under an alternative write $\bm y=\bm\mu_n+\bm\eta_n$ with
$\mathbb E(\bm\eta_n\mid\mathcal D_n)=\bm0$ and
$\operatorname{Cov}(\bm\eta_n\mid\mathcal D_n)=\bm\Omega_n\succ0$, evaluate
the model quantities at pseudo-true parameters
$\bm\theta_n^*=(\bm\beta_n^*,\sigma_n^*,\bm D_{1,n}^*,\ldots,\bm D_{L,n}^*)$,
any design-measurable choice with $\sigma_n^*>0$ and
$\bm D_{l,n}^*\succeq0$, denote them by stars, and define the mean function
$m_n^a(t)=n^{-1}\mathbb E\{\bm\sigma_\lambda(\bm y^{a*},t)^{\mathsf T}
\bm e^{a*}\mid\mathcal D_n\}$, the expected accumulated mark per observation.
The starred parameters provide an oracle reference under misspecification;
they need not generate the true data distribution. With the normalisation
of \eqref{eq:processes-main}, the oracle process has expectation
$\sqrt n\,m_n^a(t)$. Thus a mean function bounded away from zero describes
residual contributions that accumulate systematically instead of cancelling.
Theorem~\ref{thm:consistency-main} turns such a drift into power when the
primary process follows this oracle behaviour.
A vanishing mean function is a statement about the centring of the pooled
mark--ordering pair; it determines neither the covariance of a process nor
the law of its secondary processes (Section~\ref{S-sec:si-detection}).

For the projected processes the mean function is computed in
\citet[Proposition~9]{asymptoticcalibration}. We use three of its
consequences. First, a projected process accumulates the projected bias
$\bm R(\bm\mu_n-\bm X\bm\beta_n^*)/\sigma_n^*$, so a mean error inside
$\operatorname{col}[\bm X,\bm Z]$ produces no drift. Second, a covariance
error with a correct mean enters only through the covariances between
each mark and its ordering value, and these sum to zero over the
observations. It produces a drift only when the ordering values differ in
law. This is how a missing random slope is detected: the ordering must
vary with the covariate whose slope is missing, as it does when that
covariate has a fixed effect or a random slope for another factor. Third,
$O_\varnothing$ has a deterministic oracle ordering, so its mean function
shows mean errors only. The whitened process needs its own statement.

\begin{proposition}[Mean function of the whitened process]
\label{prop:mean-main}
Suppose $\bm\eta_n\mid\mathcal D_n\sim N(\bm0,\bm\Omega_n)$ and let
$\bm w^F=(\sigma_n^*)^{-1}(\bm V_n^*)^{-1/2}(\bm\mu_n-\bm X\bm\beta_n^*)$
be the whitened bias. Then, for every $\bm\Omega_n$,
\[
 m_n^F(t)=\frac1n\sum_iw_i^F\,
 \sigma_\lambda\bigl((\bm X\bm\beta_n^*)_i-t\bigr).
\]
The mean function of $F$ is the partial-sum process of the whitened bias
along the fixed prediction. It vanishes when
$\bm\mu_n=\bm X\bm\beta_n^*$.
\end{proposition}

The whitened bias includes mean errors inside the fitted span, which the
projected processes cannot show. Under rotations that part belongs to the
offset and cannot produce a rejection (Section~\ref{sec:fixed}).
Section~\ref{S-sec:si-detection} derives the mean functions of all
processes from Stein's identity. The next result bounds the power of the
exact tests for every sample.

\begin{theorem}[Power of the exact tests]\label{thm:consistency-main}
Let $T$ be the KS or the root CvM statistic of a process $a$ at one scale,
computed from an ML or REML fit, and let
$P_c(t)=n^{-1/2}\bm\sigma_\lambda(\widehat{\bm y}^a,t)^{\mathsf T}\bm c_a$
be the path of the offset alone, which is zero for a projected process. Put
\[
 \tau_n=n^{-1/2}\{T(\widehat W_n^a)-T(P_c)\}_+,\qquad
 C_{\rm CvM}=1,\qquad C_{\rm KS}=2.
\]
For every $n$, every design with $r>0$, every $B$ and every response
$\bm y$ with $\bm R\bm y\ne\bm0$, the rotations
\eqref{eq:secondary-exact-main} satisfy
\begin{equation}
 \Pr\{\widehat p>1/(B+1)\mid\bm y\}\le\frac{B\,C_T}{r\,\tau_n^2}.
 \label{eq:power-bound-main}
\end{equation}
Hence $\Pr(\widehat p\le\alpha)\to1$ for every fixed $B$ with
$\alpha(B+1)\ge1$ along any sequence of data distributions for which
$r\tau_n^2\to\infty$ in probability, in particular when $r\to\infty$ and
$\tau_n\ge\delta$ with probability tending to one for some $\delta>0$.
For the minimum-$p$ rule over $J$ columns it suffices that one column has
$r\tau_n^2\to\infty$ and that $J\le\alpha(B+1)$.
\end{theorem}

No model for the data enters: given $\bm y$, the secondary statistics are
functions of uniform directions, their mean square beyond the offset path
is at most $C_Td^2$, and $d^2\le n/r$ because the profiled error variance
is at least $\|\bm R\bm y\|^2/n$ by \eqref{eq:likelihood-split}
(Theorem~\ref{S-thm:exact-power}). For a projected process $\tau_n$ is the
height of the primary path divided by $\sqrt n$; it stays away from zero
when the mean function does and the primary process follows its oracle
version, so the exact projected tests are then consistent as soon as the
residual dimension grows (Corollary~\ref{S-cor:exact-consistency}). For
$F_E$ only the excess over the offset path counts: a departure inside the
fitted span raises the offset alone, in agreement with
Corollary~\ref{cor:blind-main}. The constants are not sharp; the bound
serves through its rate. Gaussian draws need further
conditions on the secondary processes (Theorem~\ref{S-thm:power}).

\subsection{Covariance departures without mean drift}
A covariance error need not produce a drift. Inside the fitted design
span it is invisible to the projected processes, and by
Corollary~\ref{cor:blind-main} their exact test keeps its level under it:
blindness here means calibration, not detection. Outside the span a
covariance error changes the law of the projected residual, and whether
an ordering shows it depends on the design. An omitted random slope with
a large variance makes $F_E$ and $O_\varnothing$ reject although the
fixed mean is correct. An omitted crossed factor whose covariate varies
within its groups is spread over every ordering and missed
(Section~\ref{sec:sim-detection}). These limits of single orderings are
the reason for
reading the processes as a family.

\section{Simulation study}
\label{sec:simulations}

\subsection{Design}
\label{sec:sim-design}
The main study covers five designs: single-level, nested, crossed,
five-centre and three-level. Table~\ref{S-tab:designs} gives their group
counts and random structures. Responses have fixed mean
$1+x_1+x_2+\beta_qx_1^2$, independent uniform $(0,1)$ covariates, Gaussian
random coefficients and independent standard Gaussian errors. Random
intercepts have standard deviation 1 (0.7 for three-level items), and
slopes 0.75. The fitted covariance is unstructured within each grouping
factor. Random effects are Gaussian in the main study; by
Corollary~\ref{cor:blind-main} the null results of the projected tests do
not depend on that choice. Subjects are grouping factor 1, with the
ordering $O_1$; centres or items are factor 2, with $O_2$. The five
correct-model cases and 23 alternatives omit a quadratic mean term
($\beta_q=1$), random slopes, an entire grouping factor, or both mean and
random terms. In the figures and tables, `slope 1' and `slope 2' denote
an omitted random slope of factor 1 or 2, `level 2' an omitted factor 2
and `$x_1^2$' the omitted quadratic term. Each of these 28 scenarios uses
1000 ML replicates at $n=250$, $1000$ and $4000$, with $B=999$, indicator
weights, the five scales and the minimum-$p$ rule. Every test is the
exact test of Theorem~\ref{thm:exact-main}, with $\bm c_a=\bm0$ for the
projected processes and $\bm c_a=\bm c_F$ for $F$, computed from the same
rotations. Fits use \texttt{MixedModels.jl} with symmetric whitening for
$F$. Section~\ref{S-sec:si-simulations} gives the complete protocol.

\subsection{Calibration of the exact tests}
\label{sec:sim-calibration}
\IfFileExists{tables/size-summary.tex}{\begin{table}[!htbp]
\caption{Null rejection rates of the exact tests at $\alpha=0.05$: minimum / median / maximum over process, sample-size, statistic and scale-rule cells (KS and CvM pooled; $F$, $O$ and the factor-specific processes; $n=250, 1000, 4000$; 1000 replicates per cell, Monte Carlo standard error 0.007 at 0.05). Full results are in Tables~\ref{S-tab:exact-single}--\ref{S-tab:exact-three}.}
\label{tab:size-summary}
\centering\footnotesize\setlength{\tabcolsep}{6pt}
\begin{tabular}{@{}lccc@{}}\toprule
Design & Indicator & $c=1$ & Min-$p$ \\\midrule
single-level & 0.040 / 0.054 / 0.058 & 0.043 / 0.052 / 0.061 & 0.040 / 0.051 / 0.061 \\
nested & 0.039 / 0.046 / 0.062 & 0.033 / 0.048 / 0.060 & 0.030 / 0.048 / 0.060 \\
crossed & 0.040 / 0.050 / 0.063 & 0.044 / 0.051 / 0.060 & 0.040 / 0.048 / 0.069 \\
five-centre & 0.038 / 0.049 / 0.060 & 0.038 / 0.045 / 0.061 & 0.036 / 0.048 / 0.062 \\
three-level & 0.039 / 0.050 / 0.062 & 0.034 / 0.053 / 0.064 & 0.031 / 0.052 / 0.060 \\
\bottomrule\end{tabular}\end{table}
}{}

The exact tests give rejection rates between 0.030 and 0.069 over the 360
cells of the main grid (five designs, every process, three sample sizes,
two statistics and three scale rules; Tables~\ref{S-tab:exact-single}--\ref{S-tab:exact-three}),
with median 0.050 over all cells (Table~\ref{tab:size-summary} gives the ranges by design); the Monte Carlo standard
error at 0.05 is 0.007 for an individual cell. Singular fits are retained;
they reach 66\% under the null at $n=250$ in the nested design and 65\% in the three-level one; a zero fitted variance simply removes that factor's
contribution to the ordering. The exactness is a theorem, so the table is
a check of the implementation rather than a proof of the method; the
conditional size bound does not depend on $n$.
A separate numerical check rotates and refits 200 small Gaussian data
sets, including residual rank $r=3$ and boundary covariance fits
(Section~\ref{S-sec:si-rotation-check}). The conditioning summaries and
fixed-parameter residual quadratic are preserved to $2.2\times10^{-15}$
in relative error. Reoptimised fitted orderings differ by at most
$1.5\times10^{-4}$ on the same relative scale, documenting numerical
optimisation error separately from the exact invariant-fit argument.

\subsection{Small residual dimension}
\label{sec:sim-small-r}
A separate study follows the calibrations of
Section~\ref{sec:gaussian-approx} as $r$ grows, in three situations with
one grouping factor and Gaussian covariates, random effects and errors.
\emph{Short series}: three observations per subject, a random intercept
and slope and $p=2$, so that $r$ equals the number of subjects, from 6 to
400. \emph{Twenty fixed effects}: five observations per subject, a random
intercept and $p=20$, with $n$ from 40 to 1600. \emph{Fixed effects in
proportion}: the same with $p=n/5$ and $n$ from 25 to 800, so that $r/n$
stays near 0.6. Each of the 20 cells has 10\,000 replicates under the null
hypothesis and under an omitted quadratic mean term, with ML and REML fits
and $B=999$. Three calibrations are computed from the same data and the
same Gaussian vectors: rotations; Gaussian draws with the marks divided by
the residual standard deviation $\|\bm R\bm y\|/\sqrt r$, as in
\citet{asymptoticcalibration}; and Gaussian draws with the marks divided
by $\widehat\sigma$, as in \eqref{eq:secondary-main}. We report CvM with
the minimum-$p$ rule for $O$ and ML fits unless stated otherwise;
Section~\ref{S-sec:si-small-r} gives the protocol and the results for
$O_\varnothing$, KS, indicator weights and the other levels.

\begin{figure}[!tb]
\centering
\includegraphics[width=\textwidth]{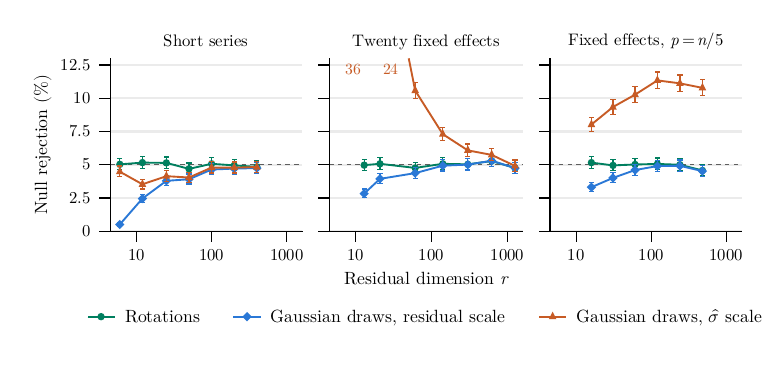}
\caption{Null rejection rates at the 5\% level against the residual
dimension $r$ (log scale) in the three situations of
Section~\ref{sec:sim-small-r}, for $O$, CvM with the minimum-$p$ rule and
ML fits: rotations, Gaussian draws with the marks divided by the residual
standard deviation, and Gaussian draws with the marks divided by
$\widehat\sigma$. The three calibrations use the same data and Gaussian
vectors; $B=999$ and 10\,000 replicates per point. Bars are pointwise 95\%
Wilson Monte Carlo intervals and the dashed line is 5\%. The two rates
printed at the top of the middle panel, 36\% at $r=13$ and 24\% at $r=21$,
lie above the plotted range (Table~\ref{S-tab:small-r-null}).}
\label{fig:small-r}
\end{figure}

Figure~\ref{fig:small-r} shows the null rejection rates. The rotations
reject between 4.6\% and 5.3\% in every cell, including $r=6$, where 67\%
of the ML fits are singular. Gaussian draws at the residual scale are
conservative by an amount that depends mainly on $r$.
They reject 0.5\% at $r=6$, 2.5--2.8\% at $r=12$ and $13$, 3.3--4.0\% for
$r$ between 16 and 50 and 4.4--4.6\% at $r=61$ and $100$, and they are
within 0.25 percentage points of the rotations from $r=121$. The scale
$\widehat\sigma$ of an ML fit acts in the opposite direction when there
are many fixed effects. With $p=20$ the rejection rate is 36\% at $n=40$,
where the mean of $d$ is 1.53, 11\% at $n=100$ and 5.7\% at $n=800$. With
$p=n/5$ it stays between 8\% and 11\% at every $n$, and the mean of $d$
between 1.11 and 1.15: this error does not vanish as the sample grows. A
REML fit removes it, with rates within 0.3 percentage points of those at
the residual scale in both situations. In the short series the mean of
$d$ is within 0.02 of one from $r=12$, and Gaussian draws at the scale
$\widehat\sigma$ are conservative as well (3.5--4.1\% for $r$ from 12 to
50).

Conservative draws lose power. Against the omitted quadratic term the
rotations reject 6.4\% of the short-series samples at $r=6$ and Gaussian
draws at the residual scale 0.8\%. The difference is 3.3 percentage points
at $r=12$, 1.9 at $r=25$ and 1.3 at $r=50$, and between 2.1 and 2.7 in the
two situations with many fixed effects when $r\le21$, with standard errors
of at most 0.2. From $r=121$ it is at most 0.5 percentage points in all
three situations (Table~\ref{S-tab:small-r-power}). A rotation is a
Gaussian draw rescaled to the observed length, so the exact test requires
no additional computation.

\subsection{Detection and localisation}
\label{sec:sim-detection}
\begin{figure}[!tp]
\centering
\includegraphics[width=\textwidth]{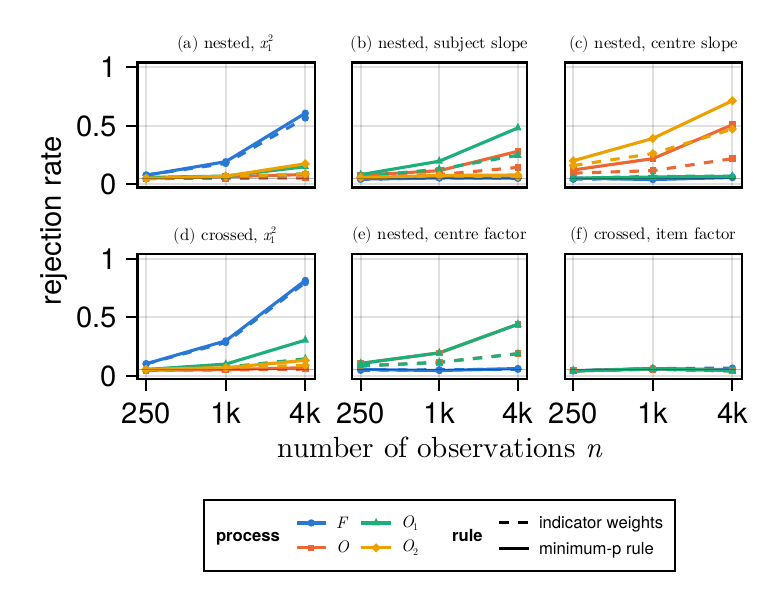}
\caption{CvM rejection rates of the exact tests against $n$ (250, 1000 and
4000, log scale) in six scenarios, each titled by the design and the omitted
term. Colours and markers identify the processes; dashed lines use
indicator weights and solid lines the minimum-$p$ rule over the scales.
$O_1$ indexes subjects and $O_2$ centres or items; in (e) and (f)
factor 2 is omitted from the fit, so $O$ and $O_1$ coincide. 1000 replicates
per point; the thin horizontal line is $0.05$.}
\label{fig:simulation-main}
\end{figure}

\paragraph{The fixed prediction and the offset.}
Against the omitted quadratic term the tests along the fixed prediction
respond and the others do not. At $n=4000$, $F_E$ rejects in 0.61 of the
nested and 0.82 of the crossed replicates, against 0.08 and 0.07 for $O$
(Figure~\ref{fig:simulation-main}). The fixed-prediction process
$O_\varnothing$, which differs from $F_E$ only by the offset, rejects in
0.49 and 0.83. In the other three designs, which have ten observations
per subject, the offset adds between 0.03 and 0.05
(Table~\ref{S-tab:exact-F-main}). This is the pattern that
Section~\ref{sec:fixed} explains: the offset helps with small subjects and
not in the crossed design. Under the omitted random slopes of the main
designs $F_E$ stays near its level, at most 0.065 at $n=4000$. Under more
severe covariance errors outside the fitted span it rejects a correct
fixed mean in 0.59--0.64 of the replicates
(Table~\ref{S-tab:exact-F-stress}); these are departures from its model
null.

\paragraph{Orderings along single factors.}
The factor-specific processes respond mainly along the factor with the
missing slope. At $n=4000$ in the nested design, omitting the subject
slope gives rates 0.48 for $O_1$ and 0.08 for $O_2$, while omitting the
centre slope gives 0.07 and 0.71. An omitted nested centre factor is seen
by $O$ (0.44). An omitted crossed item factor is missed by every process
at every sample size (0.044 for $O$ at $n=4000$), because the crossed
orderings spread it out.

\paragraph{The joint test.}
Reading several processes separately inflates the error rate. Under a
correct model some $O_l$ rejects in 0.07--0.12 of the replicates with two
grouping factors and in 0.12--0.16 with three, and some process among
$F$, $O$ and the $O_l$ in 0.09--0.24 (Table~\ref{S-tab:joint}). The
minimum-$p$ rule of Section~\ref{sec:pvalues} adjusts for the scales
within one process, not for the family of processes. The joint test of
Corollary~\ref{cor:joint-main} over all processes and the five scales
uses the same rotations. Its size lies between 0.035 and 0.066 in every
design and at every sample size, and at $n=4000$ it keeps most of the
power of the best single process. It rejects the omitted subject slope
in 0.35 of the nested and 0.56 of the crossed replicates ($O_1$: 0.48 and
0.69), the omitted centre or item slope in 0.60 and 0.58 ($O_2$: 0.71 and
0.70), the omitted quadratic term in 0.37 and 0.69 ($F_E$: 0.61 and
0.82), and both omitted slopes in 0.80 of the replicates of either design
(the factor-specific tests: 0.59--0.69). We recommend it as the single
level-$\alpha$ decision before the factor-specific paths are read.

\paragraph{Weak departures and targeted tests.}
Power against weak departures is limited. At $n=4000$, an omitted subject
slope of standard deviation 0.25 is essentially undetected (0.05 and
0.08); at standard deviation 0.5, $O_1$ rejects in 13\% and 28\% of
nested and crossed replicates
(Table~\ref{S-tab:author-extra-weak-slopes}). In the nested benchmark, a
likelihood-ratio test for the specified missing slope rejects in 63\% and
100\% of replicates, against 20\% and 48\% for $O_1$
(Table~\ref{S-tab:benchmark}). The residual process gains graphical and
factor-specific information at a cost in power
(Table~\ref{tab:benchmark-main}). Table~\ref{S-tab:localisation} gives the
joint behaviour of the factor-specific tests and a sequential reading,
$F_E$ first and then $O$ with the factor that rejects: under an omitted
slope the wrong factor alone rejects in at most 0.06 of the replicates of
the nested and crossed designs at $n\ge1000$.
\IfFileExists{tables/benchmark-main.tex}{\begin{table}[tb]
\centering\footnotesize\setlength{\tabcolsep}{4pt}
\caption{Rejection rates at $\alpha=0.05$ and $n=4000$ of the exact tests ($F_E$, $O_E$ and the factor-specific tests; CvM with the minimum-$p$ rule) against tests that name the alternative, on the same 1000 replicates of the nested and crossed designs (ML fits, $B=999$): the Wald $z$-test of $x_1^2$ added to the fitted model and the likelihood-ratio tests $\mathrm{LRT}_l$ of one additional correlated random slope for factor $l$ with the boundary-corrected reference, computed by the companion paper on these replicates. $^{\dagger}$: the fitted model already contains the slope of that covariate for that factor, so a slope of the other covariate is added; these entries are sizes, or the reaction of the test to a misspecification elsewhere. $n=1000$ and the details are in Table~\ref{S-tab:benchmark}.}
\label{tab:benchmark-main}
\begin{tabular}{@{}llrrrrrrr@{}}
\toprule
Design & omitted & $F_E$ & $O$ & $O_1$ & $O_2$ & Wald $x_1^2$ & $\mathrm{LRT}_1$ & $\mathrm{LRT}_2$ \\
\midrule
nested & nothing ($H_0$) & 0.056 & 0.048 & 0.050 & 0.035 & 0.053 & 0.042$^{\dagger}$ & 0.032$^{\dagger}$ \\
 & $x_1^2$ & 0.606 & 0.081 & 0.148 & 0.173 & 0.986 & 0.043$^{\dagger}$ & 0.032$^{\dagger}$ \\
 & subject slope & 0.051 & 0.280 & 0.480 & 0.077 & 0.064 & 0.997 & 0.148$^{\dagger}$ \\
 & centre slope & 0.056 & 0.507 & 0.065 & 0.713 & 0.052 & 0.801$^{\dagger}$ & 1.000 \\
\addlinespace[2pt]
crossed & nothing ($H_0$) & 0.051 & 0.051 & 0.055 & 0.058 & 0.050 & 0.032$^{\dagger}$ & 0.042$^{\dagger}$ \\
 & $x_1^2$ & 0.815 & 0.068 & 0.304 & 0.131 & 0.998 & 0.030$^{\dagger}$ & 0.043$^{\dagger}$ \\
 & subject slope & 0.059 & 0.163 & 0.691 & 0.056 & 0.061 & 1.000 & 0.041$^{\dagger}$ \\
 & item slope & 0.056 & 0.159 & 0.053 & 0.695 & 0.050 & 0.031$^{\dagger}$ & 1.000 \\
\bottomrule
\end{tabular}
\end{table}
}{}

\paragraph{Smoothing.}
Smoothing often increases the power of the projected processes, whereas a
wide window can weaken $F$. At $n=4000$, averaging over processes and
alternatives, the minimum-$p$ rule increases the mean CvM rejection rate
from 0.24 to 0.33 (KS: 0.18 to 0.30). These results motivate CvM with the
minimum-$p$ rule as the main test; the other statistics provide
sensitivity comparisons.

\paragraph{Scope beyond Gaussian errors.}\label{sec:sim-nongaussian}
The uniform conditional direction follows from Gaussian observation
errors, and without them the tests are no longer exact. Random effects of
any distribution inside the fitted design span are covered by
Corollary~\ref{cor:blind-main}; Section~\ref{S-sec:si-simulations}
reports a check with $t_3$ and exponential random effects. For
non-Gaussian observation errors, \citet{asymptoticcalibration} gives
large-sample results for the fixed-prediction ordering.

\section{Two small studies}
\label{sec:application}
Two classical data sets with 21 to 30 residual degrees of freedom show
the tests in the range where the calibrations of
Section~\ref{sec:sim-small-r} differ. Every model is fitted by ML. The
$p$-values of the residual-process tests are CvM with the minimum-$p$
rule and use $B=99\,999$ Gaussian vectors, so that their Monte Carlo
standard error is at most 0.0016 (Table~\ref{tab:small-studies});
Section~\ref{S-sec:si-small-studies} gives KS, $O_\varnothing$ and REML
fits. Both studies have one grouping
factor, so that $O_1=O$ and the joint test of
Corollary~\ref{cor:joint-main} combines $F_E$ and $O$.

\IfFileExists{tables/small-studies.tex}{\begin{table}[tb]
\centering
\caption{The two small studies: CvM $p$-values with the minimum-$p$ rule over the five scales, ML fits and $B=99\,999$ Gaussian vectors, the same for every entry of a row. Joint is the test of Corollary~\ref{cor:joint-main} over $F_E$ and $O$. The last two columns calibrate $O$ by Gaussian draws, with the marks divided by the residual standard deviation or by $\widehat\sigma$. The gasoline models have the fixed part of \citet[Section~10.1]{Venables2002} and a random intercept for each crude; the rat models have a random intercept and slope for each animal and are fitted to each treatment group separately. KS, $O_\varnothing$, REML fits and further models are in Section~\ref{S-sec:si-small-studies}.}
\label{tab:small-studies}
\footnotesize
\setlength{\tabcolsep}{4pt}
\begin{tabular}{@{}lrrrccccc@{}}
\toprule
 & & & & \multicolumn{3}{c}{Rotations} & \multicolumn{2}{c}{Gaussian draws, $O$} \\
\cmidrule(lr){5-7}\cmidrule(l){8-9}
Model & $n$ & $r$ & $d$ & Joint & $F_E$ & $O$ & Residual scale & Scale $\widehat\sigma$ \\
\midrule
\multicolumn{9}{@{}l}{\emph{Gasoline yield}} \\
\quad Yield & 32 & 21 & 1.030 & 0.008 & 0.015 & 0.006 & 0.013 & 0.010 \\
\quad Yield, end-point slopes & 32 & 12 & 0.873 & 0.041 & 0.025 & 0.106 & 0.124 & 0.215 \\
\quad Square root of the yield & 32 & 21 & 1.030 & 0.241 & 0.225 & 0.191 & 0.188 & 0.169 \\
\multicolumn{9}{@{}l}{\emph{Rat growth}} \\
\quad Linear, control & 50 & 30 & 1.000 & 0.070 & 0.050 & 0.082 & 0.086 & 0.086 \\
\quad Linear, thiouracil & 50 & 30 & 1.000 & 0.001 & 0.001 & 0.002 & 0.005 & 0.005 \\
\quad Linear, thyroxin & 35 & 21 & 1.000 & 0.001 & 0.002 & 0.0004 & 0.003 & 0.003 \\
\quad Quadratic, control & 50 & 29 & 1.017 & 0.272 & 0.163 & 0.198 & 0.219 & 0.200 \\
\quad Quadratic, thiouracil & 50 & 29 & 1.017 & 0.343 & 0.209 & 0.384 & 0.388 & 0.363 \\
\quad Quadratic, thyroxin & 35 & 20 & 1.025 & 0.194 & 0.169 & 0.136 & 0.175 & 0.151 \\
\bottomrule
\end{tabular}
\end{table}
}{}

\subsection{Gasoline yield}
\label{sec:gasoline}
\citet{Prater1956} recorded the percentage of crude oil converted to
gasoline in 32 distillation runs, with the end point of the distillation
and three properties of the crude. The properties take ten distinct
combinations, so the runs fall into ten crude oils with two to four runs
each. \citet[Section~10.1]{Venables2002} introduce linear mixed models
with these data. We take from that section the model with a random
intercept for each crude and fixed effects of the end point and of the
ASTM 10\% point of the crude. The crude properties are constant within a crude and lie in the
span of the crude indicators, so that $r=32-10-1=21$ whichever of them
enter the fixed part.

The joint test rejects this model ($p=0.008$). For $O$ the rotations give
$p=0.006$; Gaussian draws give $0.013$ at the residual scale and $0.010$
at the scale $\widehat\sigma$. The Gaussian $p$-values are
about twice the exact one, which is the conservatism that
Section~\ref{sec:sim-small-r} measures at this residual dimension.
Leaving out one run at a time keeps the exact $p$-value of $O$ between
0.001 and 0.031 and that of Gaussian draws at the residual scale between
0.007 and 0.052 ($B=9999$).

\begin{figure}[!tb]
\centering
\includegraphics[width=\textwidth]{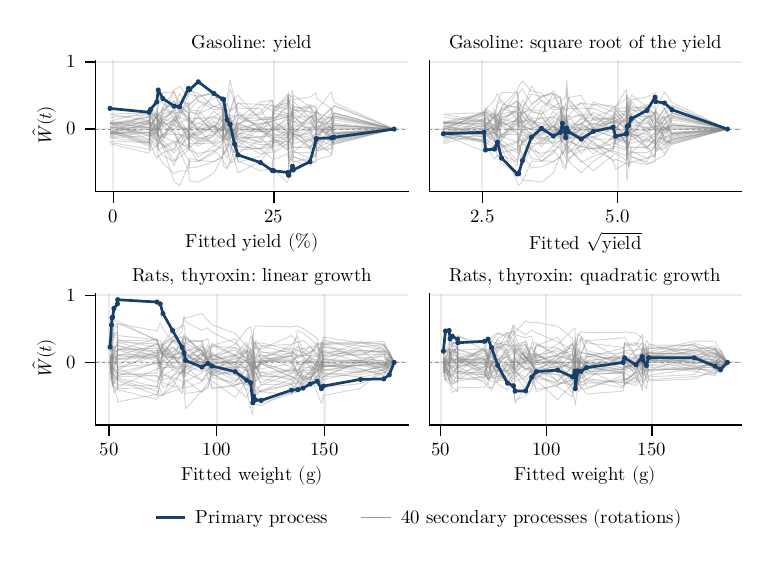}
\caption{The process $O$ with indicator weights in the two studies, for
the first model (left) and its revision (right). Dark lines join the
values of the primary process at the fitted predictions; grey lines are 40
of the secondary processes from rotations. These paths illustrate the
conditional law; the displayed collection is not an acceptance band.
Table~\ref{tab:small-studies} reports the tests.}
\label{fig:small-studies}
\end{figure}

The usual extensions of the model do not remove the departure. A random
end-point slope for each crude does not improve the fit: the
likelihood-ratio statistic is 0.003 and the fit is singular. A quadratic
end-point term has a likelihood-ratio $p$-value of 0.28. The joint test
still rejects both extended models ($p=0.041$ and $0.021$; KS $0.083$ and
$0.031$). The path of
$O$ in Figure~\ref{fig:small-studies} shows the form of the departure.
The residuals are positive at small and at large fitted yields and
negative between them, so the yield is convex in its prediction: the
departure is in the scale of the response, not in one covariate. A
squared fitted value added to the fixed part has a likelihood-ratio
$p$-value of 0.002, and the smallest fitted yield is $-0.5\%$. With the
square root of the yield as the response no test rejects (joint
$p=0.24$). The yield is a proportion, and
\citet{FerrariCribariNeto2004} use the same data to introduce beta
regression.

The model with random slopes has $r=12$ and $d=0.87$. There Gaussian
draws at the scale $\widehat\sigma$ give $p=0.215$ for $O$, against
$0.106$ from the rotations and $0.124$ at the residual scale: with $d<1$
the scale $\widehat\sigma$ errs in the conservative direction.

\subsection{Rat growth}
\label{sec:rats}
\citet{Box1950} gives the weights of 27 rats on five weekly occasions:
ten controls, ten rats with thiouracil and seven with thyroxin in their
drinking water. We use the copy distributed with \citet{Fitzmaurice2011}.
For each treatment group we fit a straight line in time with a random
intercept and slope for every animal. Every animal then leaves three
residual degrees of freedom, so that $r=30$, 30 and 21, which is the
short-series situation of Section~\ref{sec:sim-small-r}. In this balanced
design the fitted error variance equals the residual mean square: $d=1$,
and the two Gaussian scales coincide.

The joint test rejects linear growth under thiouracil and under thyroxin
($p=0.001$ in both groups) and not for the controls ($p=0.070$; KS
$0.043$). For the
seven thyroxin rats the rotations give $p=0.0004$ for $O$ and Gaussian
draws $0.003$, eight times as much; under thiouracil the two values are
$0.002$ and $0.005$. A quadratic term in time removes the rejections
(joint $p=0.27$, $0.34$ and $0.19$ for controls, thiouracil and
thyroxin). Curved growth is the expected departure here. The
likelihood-ratio test of the quadratic term, which targets it, gives
$p=0.030$, $0.0005$ and $0.0007$ in the three groups and so also detects
the weaker curvature of the controls, in line with
Section~\ref{sec:sim-detection}.

For $O$ the three calibrations lead to the same decisions at the 5\%
level in both studies. They differ in the size of the $p$-values, by
factors of two to eight where the model is rejected. The revised models were chosen after the first tests, so their
nonrejections are descriptive. Section~\ref{S-sec:si-kb07} applies the
joint test and the orderings along single factors to a crossed
response-time experiment with 1789 observations, in which $d$ is within
0.005 of one and rotations and Gaussian draws agree to within 0.01.

\section{Discussion}
\label{sec:discussion}
The inferential benefit and the limitation have the same source.
Conditioning on the part of the response inside the fitted design span
removes the random effects without estimating their distribution, which
gives exact tests with few groups, crossed factors and singular fits. It
also removes any mean or covariance departure inside that span. Outside
it, the ordering determines which departures are visible, and a targeted
likelihood-ratio test can be much more powerful when an alternative is
specified. An exact $p$-value is therefore not an omnibus certification
of a mixed model, and a rejection along one ordering does not attribute
the departure to one grouping factor. Model revision guided by these
paths is exploratory unless a subsequent selection procedure is
separately justified.

Rotations and Gaussian draws use the same random vectors and differ by a
radial factor. When the residual dimension is large and the estimated
error variance is close to the residual mean square, as in the
response-time experiment of Section~\ref{S-sec:si-kb07}, the two agree,
and the large-sample theory of \citet{asymptoticcalibration} applies. The exact calibration matters in
small samples and, more generally, when few residual degrees of freedom
remain. In the simulations Gaussian draws at the residual scale were
conservative below a residual dimension of about 100, markedly so below
25, and lost up to 6 percentage points of power; with an ML variance
estimate and many fixed effects they exceeded the level by an amount that
shrinks only as the share of fixed effects in the sample does. From a
residual dimension of about 120 the null rejection rates at the residual
scale agreed with those of the rotations to within a quarter of a
percentage point, and the power to within half a point. In the two
small studies the $p$-values of Gaussian draws were two to eight times
the exact ones. A rotation costs
no more than a Gaussian draw. Its exactness requires Gaussian observation
errors.

The fitted offset explains why tests along the same fixed prediction can
differ in power, and the bound in Theorem~\ref{thm:consistency-main}
expresses this through the observed excess above the offset. The rotation
tests complement tests of random-effect distributions
\citep{Verbeke2013,Drikvandi2017}, graphical diagnostics
\citep{Nobre2007,Loy2017,Schutzenmeister2012} and tests of specified
variance components \citep{CrainiceanuRuppert2004,Scheipl2008}.
The proofs are formalised in Lean (Section~\ref{S-sec:si-lean}).

\backmatter

\bmhead{Supplementary information}
Online Resource 1 contains the proofs of the exact results
(Section~\ref{S-sec:si-exact}), the setting of the asymptotic statements
(Section~\ref{S-sec:si-conditions}), the behaviour under alternatives
(Section~\ref{S-sec:si-detection}),
the auxiliary lemmas (Section~\ref{S-app:lemmas}), the numerical details
(Section~\ref{S-sec:si-simulations}), the details of the two small
studies and a crossed response-time application
(Section~\ref{S-sec:si-application}) and the
Lean formalisation of the proofs (Section~\ref{S-sec:si-lean}).

\bmhead{Acknowledgements}
The author thanks Nataša Kejžar and Rok Blagus, the co-authors of the
one-level paper, for discussions that shaped the processes studied here.
We acknowledge financial support from the Slovenian Research and Innovation Agency
(ARIS; programme P3-0154). We have no conflicts of interest to declare.

\section*{Statements and Declarations}
\bmhead{Funding}
This work was supported by the Slovenian Research and Innovation Agency (ARIS;
programme P3-0154).

\bmhead{Conflict of interest}
The author has no competing interests to declare that are relevant to the
content of this article.

\bmhead{Data availability}
The gasoline data of \citet{Prater1956} are distributed as \texttt{petrol}
with the R package \texttt{MASS} \citep{Venables2002}. The rat weights of
\citet{Box1950} are distributed with \citet{Fitzmaurice2011} and are
included in the repository named below. The \texttt{kb07} data set of
\citet{Kronmuller2007}, used in the supplement, is distributed with
the Julia package \texttt{MixedModels.jl}
(\url{https://github.com/JuliaStats/MixedModels.jl}, data release
\texttt{MixedModelsDatasets}, OSF project \texttt{a94tr}). The simulated data are reproducible from the seeds
recorded in the simulation code.

\bmhead{Code availability}
The repository \url{https://github.com/JakobPeterlin/exact_LMM_GOF}
contains the simulation and analysis code
in \texttt{simulations/}, including the Julia 1.12 dependency manifest,
\texttt{MixedModels.jl} 5.9 and the vendored \texttt{GOFLMM} 0.1.0
implementation. Rotations and joint tests are implemented in
\texttt{scripts/exact1\_lib.jl}; \texttt{scripts/small\_studies.jl} computes
the two small studies and \texttt{scripts/kb07\_joint.jl} the joint test of
the response-time application. The \texttt{lean/} directory contains
the Lean~4 project (toolchain v4.35.0-rc2, with the Mathlib revision pinned
in its manifest) and its check scripts. Every numbered statement of the
paper and of the Supplementary Information is stated and proved there as
a Lean theorem; Section~\ref{S-sec:si-lean} describes the formalisation
and lists the theorems.

\bmhead{Use of large language models}
Large language models were used in the preparation of this work in 2026:
OpenAI's Astra and Sol and Anthropic's Claude Opus and Claude Fable.
They assisted in drafting and editing the text, in developing and checking
the proofs, in writing the simulation and analysis code, and in writing the
Lean formalisation of the proofs described in the Supplementary Information, Section~\ref{S-sec:si-lean}; the
formal verification itself is performed by the Lean proof checker, not by a
language model, and no numerical result depends on the formalisation. The
author designed the study, reviewed all generated content, checked the
references and the numbers quoted from the simulation output and the data
analyses against their sources, and takes full responsibility for the
content of the paper.

\bmhead{Author contributions}
Single author.

\bmhead{Ethics approval}
Not applicable.

\bmhead{Consent to participate and for publication}
Not applicable.

\bmhead{Materials availability}
Not applicable.

\bibliography{refs}

\end{document}